\documentclass[useAMS,usenatbib]{article}

\usepackage{graphicx,amssymb,bm}
\usepackage{subfig,xfrac}
\usepackage{float, Abbrev}
\usepackage[fleqn]{amsmath}  
\usepackage{color,url}

\usepackage{fullpage}
\newcommand{\be}{\begin{equation}}
\newcommand{\ee}{\end{equation}}
\newcommand{\ba}{\begin{eqnarray}}
\newcommand{\ea}{\end{eqnarray}}

\newcommand{\cpg}{cm$^2$/g}
\newcommand{\sidm}{\sigma_{\rm DM}/m}

\def\simlt{\lower.5ex\hbox{$\; \buildrel < \over \sim \;$}}
\def\simgt{\lower.5ex\hbox{$\; \buildrel > \over \sim \;$}}

\usepackage{graphicx} 
\usepackage{appendix}
\begin{document}

\title{A deep-learning algorithm to disentangle self-interacting dark matter and AGN feedback models}

\author{D. Harvey$^{1}$\thanks{e-mail: {\tt david.harvey@epfl.ch}}  \\
$^{1}$Laboratoire d'astrophysique, EPFL, Observatoire de Sauverny, 1290 Versoix, Switzerland}

\date{Accepted - May 2024. Received - February 2024; in original form \today.}


\maketitle

\label{firstpage}

\begin{abstract}
\noindent  
Different models of dark matter can alter the distribution of mass in galaxy clusters in a variety of ways. However, so can uncertain astrophysical feedback mechanisms. Here we present a Machine Learning method that ``learns" how the impact of dark matter self-interactions differs from that of astrophysical feedback in order to break this degeneracy and make inferences on dark matter. 
We train a Convolutional Neural Network on images of galaxy clusters from hydro-dynamic simulations. In the idealised case our algorithm is 80\% accurate at identifying if a galaxy cluster harbours collisionless dark matter, dark matter with $\sidm=0.1$\cpg or with $\sidm=1$\cpg. Whilst we find adding X-ray emissivity maps does not improve the performance in differentiating collisional dark matter, it does improve the ability to disentangle different models of astrophysical feedback. 
We include noise to resemble data expected from Euclid and Chandra and find our model has a statistical error of $<0.01$\cpg~and that our algorithm is insensitive to shape measurement bias and photometric redshift errors. 
This method represents a new way to analyse data from upcoming telescopes that is an order of magnitude more precise and many orders faster, enabling us to explore the dark matter parameter space like never before.
\end{abstract}

\section{Introduction} \label{sec:intro}

Despite a body of evidence that supports the existence of some unobservable `dark matter', that constitutes $\sim 85\%$ of all the matter in the Universe, we are yet to truly understand the source of the gravitational anomalies we observe. 

The simplest explanation is that dark matter is a massive particle that interacts weakly with the Standard Model. With this model we are able to reproduce the observed distribution of galaxies and their observed cosmological shear \cite{DES_Y3,kids-1000,HSC_Y3,DES-KIDS}. Models of modified gravity that aim to reproduce the same observations and more, including rotation curves of galaxies \cite{MOND,teves}, often fail to reproduce the observed offsets between the gravitational lensing signal and the dominant baryonic component in merging galaxy clusters \cite{separation,minibullet,cannibal}. Meaning that dark matter has become a pillar of the cosmological model.

The success of the cold and collisionless dark matter paradigm, although enabling precise cosmology, has also presented very little information on what exactly it is. As such, in a bid to understand more about dark matter we continue to search for deviations from this simplistic model. Indeed if we look at the small scales we begin to observe discrepancies. For example, studies suggest that there is an unexplained anti-correlation between the density of the dwarf galaxies in our Milky Way and their peri-center \cite{Kaplinghat2019}. Subsequent work has put this significance of this finding in to question, finding that assumptions that go in to the Milky Way potential can dramatically affect this correlation \cite{anti_corr_dwarf}. However, that being said, the anti-correlation still persists in many of the models and cannot be explained with simple baryonic physics. In addition to this, the observed diversity of rotation curves of dwarf galaxies similarly cannot be explained by hydro-dynamical simulations, even if the feedback model is altered \cite{SIDM_diversity, galaxy_diversity}. One solution to both these discrepancies is that dark matter has some finite { \it  self}-interaction on the scale of $>1$\cpg~on dwarf galaxy scales.

Self-interactions are common and expected in the natural world. Although terrestrial experiments have heavily constrained the dark matter - standard model interaction (see \cite{direct_det_rev} for a review), the self-interaction remains relatively unconstrained, with cross-sections of $\sidm>0.1$\cpg~still consistent with data. Recent studies of self-interactions in dwarf galaxies, where gravo-thermal collapse can induce a wide variety of diverse rotation curves, estimate the cross-section to be $\sidm\sim100$\cpg~\cite{Correa2020}, whereas constraints on cluster scales place upper limits closer to $\sidm<0.2$\cpg~\cite{velDepCross,SIDM_BAHAMAS}. However, despite this tentative detection of self-interacting dark matter (SIDM), they are based on limited observations of classical and ultra faint dwarf galaxies in our local group. If we are to test these theories we need ``smoking-gun" detection on galaxy clusters scales.

Galaxy clusters have long been ideal laboratories to study dark matter. Their massive concentrations can heighten any subtle change in the particle physics model, altering formation and structure at an observable scale \cite{SIDMModel,SIDMSimA,SIDMSimB,SIDMSim,RobertsonBAHAMAS}. Moreover, their abundance means that dark matter self-interaction can be studied statistically over many objects using gravitational lensing that does not assume any dynamical state \cite{Harvey15}. Many observables for dark matter have been proposed in the last decade, from measuring shapes \cite{SIDMTest,reconciling,intrinsicAlignmentSIDM}, the number of strong lensing arcs \cite{SIDMCore,RobertsonBAHAMAS}, Brightest Cluster Galaxy wobbles \cite{darkgiants,Harvey_BCG}, trails in dark matter halos \cite{dmtrails_2,Harvey_trails}, peak offsets in merging clusters \cite{SIDMModel,Harvey14,HARVEY13,bulletcluster}, and mass loss in merging clusters \cite{impactpars}. However they are often equally sensitive to baryonic feedback \cite{shapes_clusters_sidm,SIDM_BAHAMAS} making them degenerate with collisionless dark matter. 

It is clear by the numerous studies looking at different aspects of the relaxed and merging clusters that self-interactions have a non-trivial impact on the distribution of dark matter in galaxy clusters. In a bid to find a generalised method that encapsulates all these different observables we use classical image techniques from Machine Learning to simply learn what differs self-interactions from collisionless dark matter. The advantages of this method is that we can input not only different dark matter models but also different models of baryonic feedback, such that the network can learn to disentangle those key signals that differentiate the two. Moreover, in the coming year there will be a deluge of data from space telescopes such as Euclid \cite{EUCLID} and SuperBIT, a balloon borne telescope \cite{superbit}. Individually modelling all the clusters in these large samples will quickly become unfeasible. With a pre-trained intelligent model, we are able to efficiently exploit the entire datasets of thousands of clusters whilst forward modelling systematics associated with both the theory and observations.

Machine Learning (ML) in astronomy is not a new concept and has been used in cosmology for many different purposes including parameter inference or as a tool to speed up expensive computational tasks. At the heart of ML is a Neural Net (NN), which is a combination of linear mathematical operations or nodes to imitate a non-linear equation. By taking some input data, the goal of the NN is to change the weights of each node until it can reproduce some output or ground truth as best as possible, where``best as possible" is defined by a cost function input by the user, often the mean-square error. 

A classic application of NN in astronomy is to used supervised learning (i.e. all inputs are paired with a labeled output) to infer a cosmological model from some input data. For example, various studies have used a special form of NN called a Convolutional Neural Network (hereafter CNN, for more see Section \ref{sec:method}) that can take two-dimensional image data and output a labelled classification of modified gravity \cite{mertenMachine,darkAI}. 


 The goal of this paper is build a machine that can efficiently differentiate between different models of dark matter using observationally realistic inputs in a non-parametric way. This way, when it is fed real data we can make inferences on what dark matter is. To this end we develop a machine learning algorithm, based on a supervised, CNN that is able to classify between different self-interacting dark matter models. We use hydro-dynamical simulations of galaxy clusters to train our models and validate them on data that will be similar to what we will observe. We have specifically structured the paper to read as fluidly as possible. To this end we have put many of our tests in the supplementary information, that although important to understanding the behaviour of the model, may not be of interest to the general reader and hence we keep the main results here in the body. Specifically, Section \ref{sec:method} outlines the basics behind what a Convolutional Neural Network is, we then present the data we use through out the paper, and finally the more detailed overview of the network including the specific architecture. In Section \ref{sec:results} we outline the results of tests including the accuracy of our model and how these can be interpreted beyond discrete classifications. We then present our observational pipeline in Section \ref{sec:obs} that converts idealised simulations to noisy observations and present our final predicted constraints. In Section \ref{sec:disc} we discuss and present our conclusions in Section \ref{sec:conc}.

\section{Method}\label{sec:method}
A neural network's (NN) ability to agnostically learn patterns in data makes them an extremely attractive method to probe astronomical data where data volume is large and patterns can be subtle. That being said, a NN requires one-dimensional data as an input and therefore any two-dimensional image data must be reshaped or ``flattened'', resulting in a large number of inputs and hence a very large NN with many millions of parameters. Furthermore, flattening the image removes any locality information (i.e. pixels do not care where they are in the image) that is often important. Adding convolutional layers to a neural network prior to flattening can circumvent these issues. Firstly, when applying a convolution to the 2D image, the values of the convolutional kernel are the inputs to the NN, not the direct pixel values. This drastically reduces the parameter space. Moreover, convolutions can reduce the size of the image meaning that when the image is finally flattened, the size of the network isn't overly large, and hence many layers can be added resulting in a ``deep" network \cite{deep_cnn,deep_focal_loss,deep_residual_learning}. Secondly, the convolution maintains locality, meaning that pixels near each other can exchange information. This makes CNN's extremely powerful in image analysis.

In order to efficiently optimise the NN (i.e. the weights associated with each mathematical operation of a deep NN), the gradient of the weight with respect to the final cost function needs to be understood. This can be solved via ``back-propagation", which calculates this from the final node, backwards through the NN providing an analytical method to optimisation. In the past this would require some in-depth programming, however with the development of widely available, easy-to-use tools such as {\tt Tensorflow} (which is used throughout this study), important aspects of machine learning like this have been standardised, making ML much more accessible and bench-marked.

In this section we take the reader through the two key ingredients of a CNN: 1.) the input data on which the CNN is optimised and 2.) the architecture (how the network is structured including how many convolutions, and what mathematical operations are carried out) of the CNN.

\subsection{Data - hydrodynamical simulations} \label{sec:sims}

A machine learning model is only as good as the data it is fed. To this end we adopt the BAHAMAS-SIDM, hydro-dynamical simulations that include self-interactions of dark matter and a full prescription of baryonic feedback \cite{BAHAMAS}. Each model is simulated with a cosmological box-size of $400$ Mpc$/h$, with $2\times1024^3$ particles and the WMAP9 cosmology \cite{WMAP9}. The dark matter particle mass is $m_{\rm DM}=5.5\times10^9M_\odot$ and the (initial) gas particle mass is $m_{\rm gas}=1.1\times10^9M_\odot$. Below z=3, the Plummer-equivalent gravitational softening length is 5.7kpc in physical coordinates. 

{\tt BAHAMAS} is run using a modified version of {\tt Gadget-3} \cite{gadget2} and include most known-sources of feedback calibrated for the OWLS project \cite{OWLS} including subgrid models of radiative cooling \cite{OWLScoolingRates}, feedback from active galactic nuclei (AGN) \cite{OWLSagn}, star formation \cite{OWLSstarFormation}, and stellar and chemo-dynamics \cite{OWLSstellarEvoChemo}. In order to match the known fraction of gas in clusters, the stellar and AGN feedback parameters have been fine-tuned \cite{BAHAMAS,BAHAMASB}.

The dark matter self-interactions are elastic, velocity-independent and isotropic. At each time-step the nearest particles (using a fixed radius) are found and using a probabilistic recipe, each particle is scattered, conserving momentum and energy. For more please see \cite{RobertsonBAHAMAS}.

We simulate three models of collisionless dark matter, each with a differing level of Active Galactic Feedback level (whereby there is uncertainty in the fine-tuning with differing amounts of feedback from AGN still consistent the observed gas mass fraction), and three self-interacting dark matter models, $\sidm=0.1,0.3,1.0$\cpg. 
Finally we take one velocity-dependent cross-section (vdSIDM), whereby the cross-section follows:
\be
\sigma_{\rm DM}/m = \sigma_{0}\left(1+\frac{v^2}{w^2}\right)^{-1},
\ee
where v is the relative particle velocity and can be approximated as $v=\sqrt{GM_{\rm vir}/r_{\rm vir}}$. $\sigma_{0}$ is the normalisation set to be $\sigma_{0}=3.04$\cpg, and $w$ is the turn over velocity which itself is related to the ratio of the mass of a force mediator to the dark matter particle mass. Here we take $w=560$km/s. For more on this model please see \cite{RobertsonBAHAMAS}.

Each cosmological box is simulated down to a redshift of 0 whereby we
 extract the 300 most massive clusters from four redshift slices: $z=0., 0.125, 0.250, 0.375$, resulting in a total of 1200 clusters per model. We cut-out a cuboid around the cluster of size $2\times2\times10$Mpc and project along the larger $z$-axis. Given that this is a path finding project we decide to keep the training set tractable and not project on to the other two axes. This would enlarge the set and make training models computationally expensive. We test to see the impact this has on the model and how much extra information there maybe in increasing the training set size. We find that any information gained in increasing the training set would be less than 2\% (for more please see supplementary information \ref{app:trainingset}). By projecting to 10Mpc we account for most of environmental structure associated with the cluster. We then normalise each mass map such that the values lie between 0 and 1 (i.e. dividing by the maximum value in the map). We test different normalisations including dividing by the critical lensing convergence  ( assuming some source redshift of the cluster, see equation \ref{eqn:convergence} ) or the mean of the map and find that the performance is not dependent on the normalisation. For each cluster we extract two-dimensional maps of the total matter, the stellar matter and the X-ray emissivity (see \cite{BAHAMASxray} for more).
 
\begin{figure}
\centering
\includegraphics[width=0.5\linewidth]{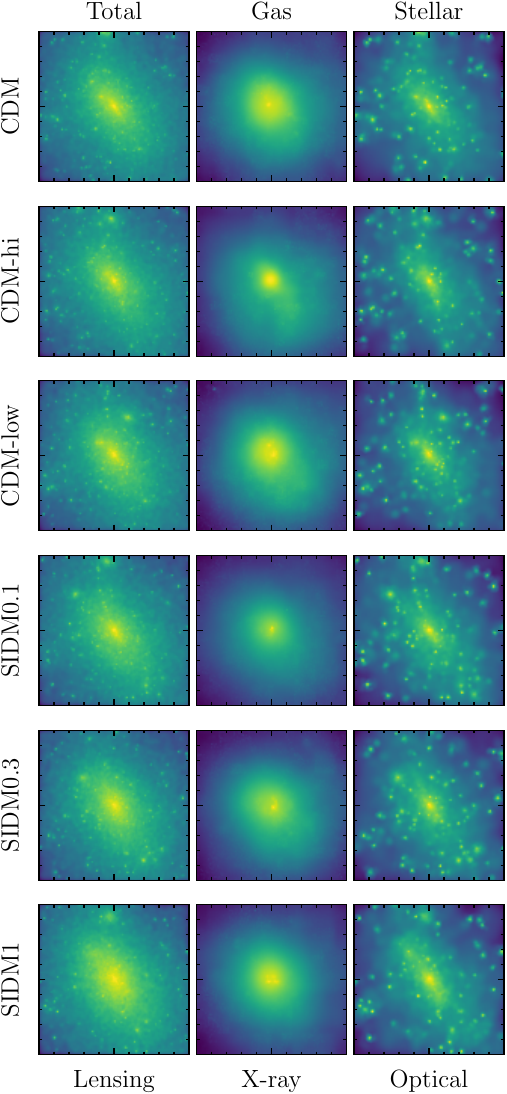}
\caption{ An example of a simulated cluster at a redshift of $z=0$, simulated in six different cosmological models. The $y$-axis label states the model used, the ``low" and``high" corresponds to level of AGN feedback. Each SIDM is simulated with the fiducial AGN. The title of each column shows the type of matter and the $x$-axis label states the observable.  }
\label{fig:data_example}
\end{figure}

 We show an example simulated galaxy cluster in Figure \ref{fig:data_example} at $z=0$ in the six simulations (rows) and input channels (columns). We show the type of matter in the title of each column and the observable at the bottom of each column.

Since the computational cost to simulate cosmological boxes of self-interacting dark matter is high, we only a have a hand full of simulated cross-sections. As such, throughout this paper (unless explicitly stated), we will adopt a CNN that will attempt to classify images of galaxy clusters in to these different values of self-interacting dark matter. That is, the input will be clusters taken from cosmological, hydro-dynamical simulations, simulated at discrete self-interaction cross-sections. The target of the CNN will be to take some blind data and classify it in to one of the discretely simulated cross-sections. Only at the end of this paper do we discuss how we go from a classification problem to a regression.

 Unless otherwise stated we train our model on 80\% of the total dataset (i.e. 960 clusters per model) and use the other 20\% (240, which the model has not seen) to validate. This allows us to test for over-fitting. Throughout this paper we will only ever train on two models of SIDM (SIDM0.1 and SIDM1), leaving the vdSIDM and SIDM0.3 purely for validation at the end. We also generate tests data for CDM with a fiducial AGN on three extra random seeds at a redshift of $z=0.25$ for more independent testing.


\subsection{CNN Architecture}\label{sec:arch}

How the Neural Net is set up including the order of convolutions to the image, how many layers to add and how and when the image is flattened is known as the structure or architecture of the CNN and can have a big impact on the performance. In an ideal work, we iterate over an infinite number of different architectures and find the one that does the best. In practice we test a variety of architectures spanning a range of complexity, from a very simple CNN consisting of 5 layers with $1,203,747$ free parameters; BIPARE, a model taken from \cite{mertenMachine}, consisting of $4,185,156$ parameters and finally Inception also from \cite{mertenMachine} consisting of $13,462,756$ parameters. We also discuss how these are altered to allow observationally matched data products later in the paper (see Section \ref{sec:obs}).

\subsubsection{ Input Channels - Observables }
Formally CNNs were developed to parse RGB images such that the input would be a three dimensional box consisting of NxNxM pixels where N is the number of image pixels and M is the number of colours. The CNN then simply adds these layers together before beginning the first convolutional layer (with the weights of each input optimised). We imitate a RGB image here by including the three main observables of a galaxy cluster: total mass from gravitational lensing, the stellar mass for optical imaging and the X-ray emission (for more on how these are derived, please see \cite{BAHAMASxray}).

We extract the three observable components of the cluster and bin them to $20$kpc, representing what we would expect high-resolution space-based weak lensing maps. Each map must have the same input dimensions for these stand alone models. We show how we deal with differing data structures and observational noise later in Section  \ref{sec:obs}.

\subsubsection{ Data augmentation }
Often in scenarios where the computation of training data is expensive it is possible to augment the data by manipulating the inputs (for example rotating, zooming or changing the contrast). This is only valid as long as the physics you are probing is invariant to the augmentation. For example, changing the contrast in this case would be changing the concentration of the halo and hence would alter the physics. However, rotating the image will only provide the convolutional kernel with more aspects to look at the data.  This has proved a successful method before in astronomy with galaxy classification \cite{galaxyZooCNN}.

In a comparative way, simulating sufficiently high resolution galaxy clusters, with accurate baryonic feedback and large enough volume to garner enough massive clusters is computationally expensive. This results in a relatively low available dataset (as compared to many neural network tasks that have hundreds of thousands of data points). As such, we augment the data to increase its size. The orientations of the projected clusters have no dependence on the interactions and could have been extracted from the simulations in any coordinate system. As such we are able to rotate and flip the clusters with no impact on the physics of the cluster it-self. When rotating we use a ``reflect" method at the boundaries where data is missing.

\section{Initial Results}\label{sec:results}

In this section we show the main results of our study. We keep the focus on the scientific output and application and keep the tuning of hyper-parameters, null-tests and other non-essential tests to the supplementary information to aid the flow of the manuscript. Should the reader be interested, please see \ref{app:metadata}. In brief we find:
\begin{enumerate}
    \item Of the three architectures we tried (simple, moderate and complex), the most complex (Inception model) is the most accurate and robust for the CNN.
    \item Maps of total matter combined with stellar provides the most constraining power, with X-ray emission only contributing small amounts of information.
    \item Augmenting the data with random flips and rotations improves the performance of the model by 33\%.
    \item The learning rate that provides the best performance is LR$=10^{-3}$.
\end{enumerate}
The left hand-panel of Figure \ref{fig:model_sensitivity} shows the performance of our base model as a function of the learning epoch for the training set (blue) and the validation set (red). This model includes all three input channels and an Inception architecture. We find that the base model reaches an accuracy of $\sim80\%$. With our base-line model we can now test it in a variety of ways to ensure we understand exactly how it is behaving. In this section we ask the following questions:
\begin{enumerate}
    \item Where and how is the model picking up information to classify different dark matter models?
    \item How does the model react to different levels of AGN feedback - can it distinguish between different versions of CDM?
    \item Does the model confuse CDM with different forms of feedback and SIDM?
    \item How well does the model do when tested on completely different simulations?
\end{enumerate}

\subsection{Data Sensitivity}

We first ask - how and where is the CNN picking up information in order to classify each dark matter model? To do this we calculate the \emph{ permutation importance} of the trained inception model in two scenarios. \emph{Permutation importance} is the act of training a model then random shuffling different regions of the validation data to see how the accuracy degrades \cite{perm_importance}. If the accuracy drops significantly, then that region that was shuffled has clearly a lot of weight in how good the model is. We thus carry out two permutation importance tests. 

\begin{figure}
\centering
\includegraphics[width=0.33\linewidth]{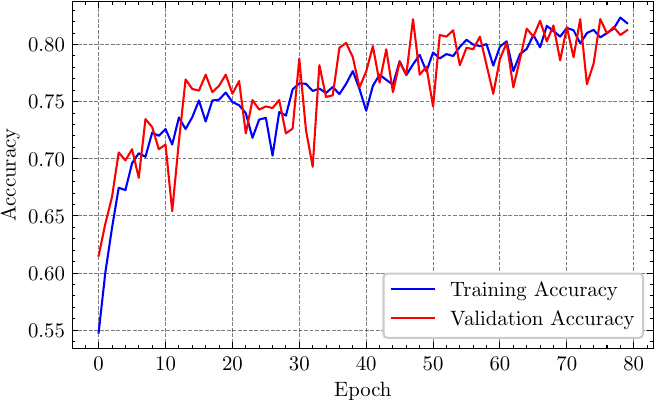}
\includegraphics[width=0.33\linewidth]{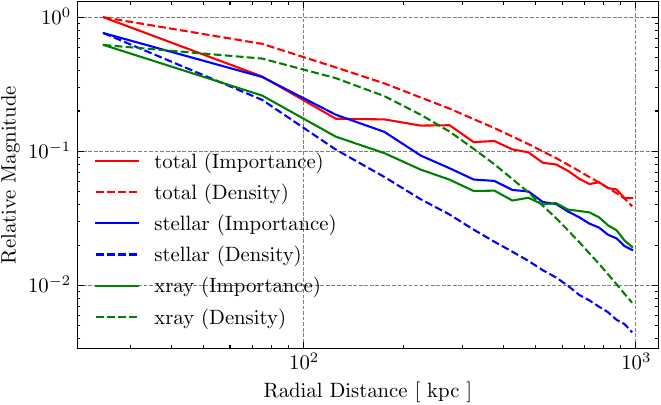}
\includegraphics[width=0.29\linewidth]{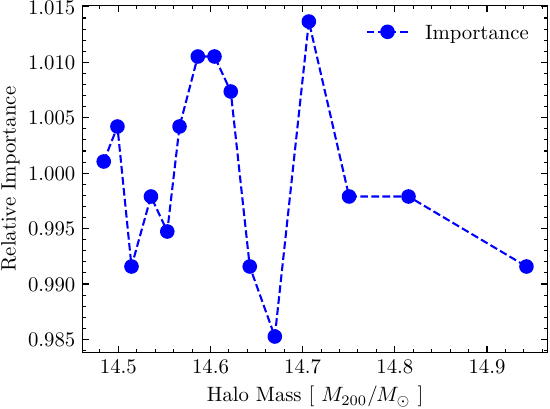}
\caption{ {\it Left:} Evolution of training for the base model (Inception plus all three input channels), classifying images of clusters in to three dark matter models ($\sidm=0,0.1,1.0$\cpg. The blue shows the training classification accuracy and the red the validation, showing how the model is not overfitting. \emph{Center} Permutation importance (PI). We calculate the relative importance of each radial bin for each input channel. We show the mean importance (solid lines) and the mean density profile (dashed line). We normalise the importance to the most important radial and input channel bin, and the density to its respective input channel. We find that the importance of each mass component falls off exponentially from the centre of the halo, with a total matter increase in importance at radial bins $>200$kpc. \emph{Right} We carry out the same PI for different mass bins normalise to the average important. We find no dependence of importance with mass.  }
\label{fig:model_sensitivity}
\end{figure}

The first is that for each input channel (total, stellar and x-ray) and radial bin with respect to the centre of the cluster, we randomly jumble the data, to remove all signal. We then calculate the accuracy of the \emph{trained} model on the new jumbled data to determine the importance of the pixels in that radial bin and channel. We define pixel importance by the ratio of the fiducial accuracy with the new accuracy normalised to the area of the radial bin. Middle panel of Figure \ref{fig:model_sensitivity} shows the results. We also calculate the normalised mass density of each input channel and see if the importance correlates with this in any way. We find that the average importance (solid lines) peaks in the center of the cluster where the density of matter is the greatest and falls off exponentially. We find that around $\sim 200$kpc the relative importance of the total matter increases with respect to the stellar matter and the X-ray emission, reflecting its matter dominance in the cluster.

The second test is to see how important halo mass is in the model. Naively we would expect larger mass halos to have more weight since they have a large difference between interacting dark matter and cold dark matter. The right hand panel of Figure \ref{fig:model_sensitivity} shows the importance of each mass bin to the model. We find that there is no trend with halo mass, meaning that it would be insensitive to any selection bias or selection mis-match between simulated halos and observationally selected halos. This is interesting since we would expect the model to perform better on larger halos where higher dark matter densities create a bigger difference between models.

\subsection{Alternate astrophysical models }
Up until now the models had been trained on three models of dark matter, collisionless, $\sidm=0.1$~and $\sidm=1.0$\cpg. However, the astrophysical processes associated with star formation, cooling and black hole feedback (i.e. the ``baryonic" model) have been fined tuned to ensure the simulations match a set of observables \cite{BAHAMAS,RobertsonBAHAMAS}. However, since the observations have error bars, the simulated data represents a single choice of parameters that produced obsedrvations that lie within some acceptable range (i.e. consistent within one-sigma of the observations). Indeed, changing the model (more specifically the amount of energy feed back by the Active Galactic Nuclei or AGN ) can alter the stellar mass distribution and the fraction of gas in the clusters (but within the error bars of the observations). Of course, if these change the models trained on a different set of data may no longer be valid. As such, we carry out a study in to the impact of differing baryonic models on the CNN. To do this we do carry out the following tests:
\begin{enumerate}
    \item How well can a CNN differentiate between different classes of CDM and what information is key in this process?
    \item Do different forms of baryonic feedback get confused with models of SIDM?
\end{enumerate}

\begin{figure*}
\centering
\begin{minipage}{\textwidth}
\includegraphics[width=0.44\linewidth]{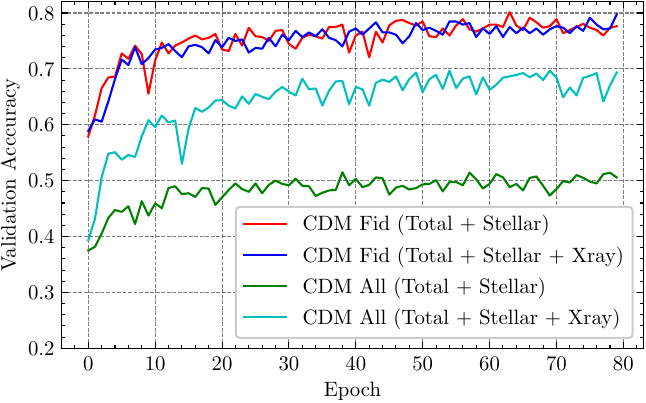}
\includegraphics[width=0.55\linewidth]{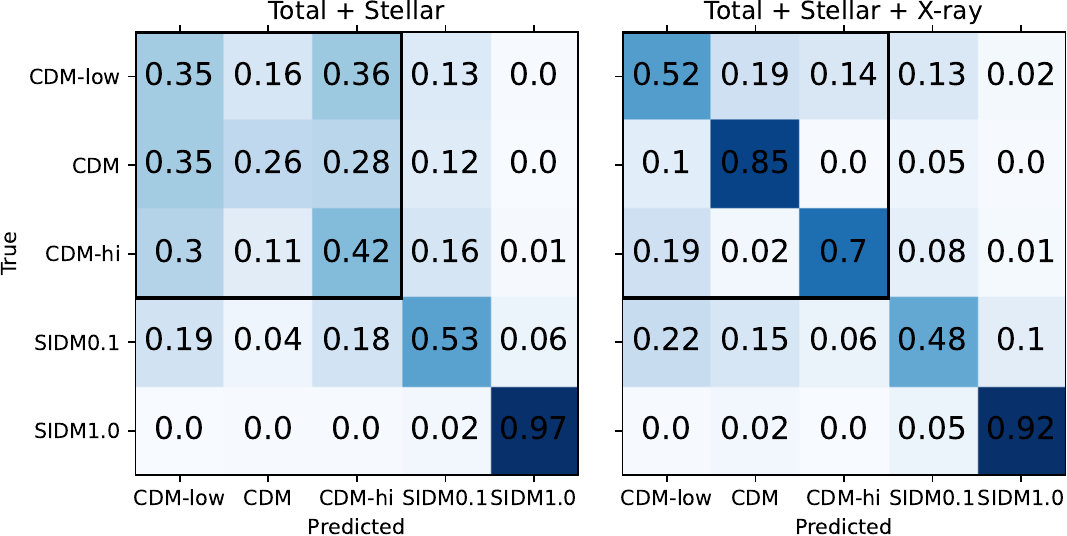}
\end{minipage}\hfill
\caption{Dependence of the CNN to different models of baryonic feedback. \emph{ Left:} We train two models with two different inputs. The first model we train has one, fiducial CDM model (plus SIDM0.1 and SIDM1) with two inputs (red, total plus stellar matter) and three inputs (total, stellar and X-ray). We find in this case adding X-ray does not improve the performance. The second model has three CDM models, low, fiducial and high AGN CDM models (plus SIDM0.1 and SIDM1 at fiducial levels), again with two (green) and three (cyan) input channels. In this case, the performance is dramatically improved with the inclusion of X-rays. \emph{ Right two matrices:} Confusion matrices for the green model (left) and the cyan model (right). We see that the improvement is purely in the CDM sector where X-ray information greatly enables the CNN to differentiate between levels of AGN feedback.  }
\label{fig:baryon_test}
\end{figure*}

\subsubsection{How well can a CNN differentiate between different classes of CDM and what information is key in this process?} 
Since the most important variable is the feedback of AGN in galaxy clusters, we focus our study on this.
Hence to answer this question we train our `simple' CNN on three models containing collisionless dark matter and three forms of AGN feedback: High, Low, and Fiducial. We then look at which input channels help differentiate between them. Figure \ref{fig:baryon_test} shows the results. The left hand panel shows the validation accuracy of two different models with two different channel inputs. The red and blue show CNN's where we classify three models: CDM, SIDM0.1 and SIDM1 using two channels - total and stellar (red) and three channels - total and stellar and X-ray (red). The cyan and green represent models where we classify five models, CDM-low AGN, CDM-hi AGN and CDM-fiducial. Interestingly, if we train the CNN with one model (fiducial) of CDM and two SIDM, adding X-ray sees no improvement in the model ( blue with and red without X-Ray information). However, if we train with three models of CDM (low, high and fiducial AGN) and two models of SIDM, adding X-ray information significantly improves the accuracy ( cyan with and green without X-Ray information). This is because the X-Ray, although not sensitive to the dark matter model, is extremely sensitive to the baryonic model used. 

\subsubsection{ Do different forms of baryonic feedback get confused with models of SIDM? }

Despite the absolute improvement in accuracy in the model, when looking more closely we find that adding X-ray information purely aids the CNN to disentangle different levels of AGN feedback, and actually increases the confusion between CDM and SIDM0.1. To do this we look beyond just accuracy and look at the confusion matrices of the model. A confusion matrix is a grid where each row represents the true input label, and each column the output prediction. So a perfect model would have a completely diagonal confusion matrix, any off diagonal elements shows how the model confuses classifications.  The right hand panel of Figure \ref{fig:baryon_test} shows the confusion matrices for the two and five classification CNNs, with the CNN using two channels (total and stellar matter) on the left and the CNN using three channels (total, stellar and X-ray) on the right. We mark out the three CDM models to help the reader. This increase confusion when including X-ray is because the model attempts to separate AGN feedback and SIDM at the same time, which have different maximum likelihood weights. Therefore if it optimises over three CDM and two SIDM it will favour its ability to disentangle CDM. Therefore, if two dark matter models have the same level of AGN feedback, it will up-weight the X-ray information since this is useful for AGN, resulting in a higher level of confusion in the dark matter sector. This clearly shows how important X-Ray information is with regard to our ability to understand AGN feedback but perhaps at the consequence of its ability to classify self-interacting dark matter models.

\subsection{Testing on different seeds}
Our final test on the fiducial model is to test the model's accuracy when validated on simulations of a different seed. The `seed' refers to the random number generator that creates the initial perturbations in the cosmological simulations. The same seed will create the same structures in the same place in a given simulation (approximately). Therefore the model may be trained on cluster A in CDM for example and then tested on the same cluster but in a different simulation so the model will have seen some part of the data before and thus the training and validation set might not be completely independent. If we simulate using a different seed it will generate entirely new and different structures and therefore act as a completely independent test and hence will be a true test of the models ability to generalise beyond its immediate training set. We begin by training ten separate models on the fiducial data with three nominal DM models. Each model draws a random subsample to train and validate on. This enables us to get an estimate in the error of the model. We then test each of these models on the new clusters from three different seeded simulations, all in a CDM scenario at a redshift of $z=0.25$. We find that the model accuracy does degrade a little as expected, with a drop in 10\% in accuracy. 


\begin{figure*}
\centering
\begin{minipage}{\textwidth}
\includegraphics[width=0.30\linewidth]{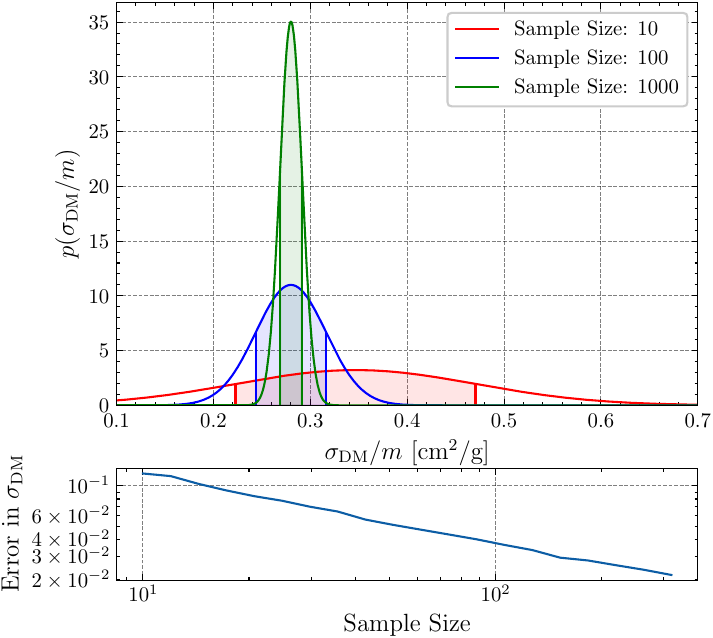}
\includegraphics[width=0.34\linewidth]{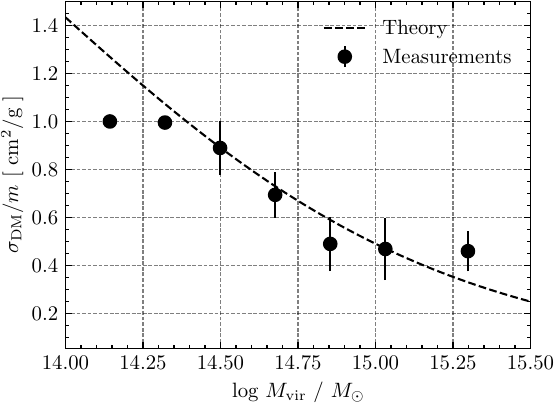}
\includegraphics[width=0.35\linewidth]{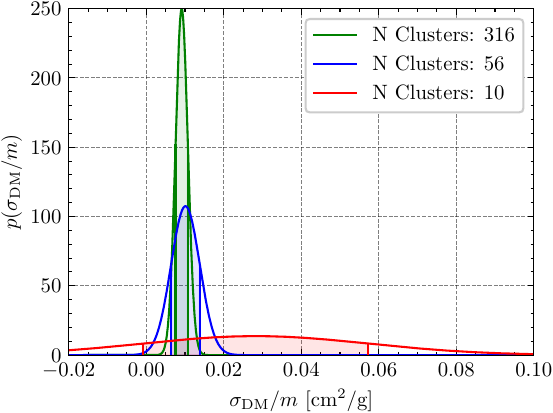}
\end{minipage}\hfill
\caption{  Combining estimates from individual clusters. We interpret the output weights of the CNN as probabilities and use them to make estimates of the cross-section for individual clusters. \emph{Left: SIDM0.3} Having trained our CNN on CDM, SIDM0.1 and SIDM1.0 we then test on the unseen SIDM0.3 and estimate the cross-section from the output weights. We run mock surveys of different sample sizes and shuffle the training sets multiple times to estimate uncertainty. We predict the estimated cross-section for the blind simulations for three sample sizes and find there is no bias. We estimate the precision as a function of the sample size in the lower plot. \emph{Center: vdSIDM} We test our model that is trained on a velocity independent cross-section on a simulation of velocity dependent SIDM. We split the test set in to mass bins and measure the cross-section. The dashed line shows the expected theoretical cross-section and the points show the model predictions.  \emph{Right: Baryonic Uncertainty}. We train on CDM-low AGN, CDM fiducial AGN and two SIDM models and test on the unseen CDM-hi AGN. We find on individual clusters we can be biased up to $\sidm=0.1$\cpg, however this is reduced with larger samples sizes, with an expected bias of $\sidm<0.02$\cpg.}
\label{fig:combine_estimates}
\end{figure*}

\section{Combining individual estimates}

In a classification problem, the final layer has the same number of nodes as there are classifications. The weights of these nodes correspond to which class the CNN thinks is the most likely, so they can be interpreted as probabilities. To this end to get an estimate of the probability for an individual cluster we interpret the output weight as a log probability, such that the conditional probability of dark matter given the model $X$ for the $i$th cluster is,
\be
p_i(\sigma_{\rm DM}/m  | {X} ) = e^{X_i}.
\ee
This gives a poorly sampled PDF of the cluster (precisely 3 points at $0$, $0.1$ and $1.0$\cpg). Since each estimate of the cross-section is the product of each PDF we find that the PDF of the cross-section for $n$ clusters is
\be
p(\sigma_{\rm DM}/m) = \prod_{i=0}^{i=n}{ p_i(\sigma)}. \label{eqn:combine}
\ee
We then find the expected cross-section from final probability distribution via,

\be
\hat{\sigma}_{\rm DM}/m = \sum{p(\sigma_{\rm DM}/m) \sigma_{\rm DM}/m}.
\ee

 To test this we train our Inception model on three dark matter models - CDM, SIDM0.1 and SIDM1.0. We then estimate the cross-section of all clusters from the SIDM0.3 ($\sidm=0.3$\cpg) model. Since the Inception model has never seen this cross-section it will be interesting to see how well it performs (although caveat that it was generated using the same seed as the previous simulations so will have seen similar clusters). The left hand panel of figure \ref{fig:combine_estimates} shows the joint estimates of the cross-section. 

To derive the error-bars we do two things:1.) Derive the epistemic uncertainty in the model due to finite size of the training data (since each training-test set split will have slightly different optimisations). To do this we Monte Carlo the data set, each time making a new random separation of the training and validation set to see what dependence the model has on the selected samples. Because each training takes some time we Monte Carlo the model twenty times. 2.)  The aleatoric uncertainty due to the finite size of the available observational dataset. We have 1200 galaxy clusters in the SIDM0.3 validation set. We simulate surveys of different sizes by splitting these in to sub-samples (as show in the legend of each figure). If there is a sub-sample size of 10 galaxy clusters we generate 11 sub-samples and look at the scatter between these. We then fit a Gaussian PDF to these distributions and show the results in the left hand panel of Figure \ref{fig:combine_estimates}. We find that for sub-sample sizes of 10, 100 and 1000 that the estimator is unbiased. We also find that for 10, 100 and 1000 clusters we would expect an error of approximately $\delta\sigma_{\rm DM}=0.15,0.05,0.01$\cpg~respectively. We calculate the expected error as a function of sample size and show it in the bottom panel of this Figure.

We finally ask the question, what happens if we apply our model trained on a velocity independent cross-section to a simulation with a velocity-\emph{dependent} cross-section. We split the test set of the vdSIDM model in to mass bins and get estimates of the cross-section for each bin using equation \eqref{eqn:combine}, with each bin containing an average of 45 clusters. The central panel of Figure \ref{fig:combine_estimates} shows the results. The black dotted line shows the theoretical cross-section and the points show the prediction from the model. We find that the model traces well the expected cross-section. When the cross-section goes above $\sidm=1$\cpg, the model breaks down since we have no simulated data here.

\subsection{ If we train on a form of baryonic feedback that isn't correct, will we infer the incorrect model of dark matter?}

Clearly, one of the largest concerns is that should the Universe not exhibit the type of star formation and feedback in our training set (such that the our simulations do not accurately reflect the true Universe) will we confuse CDM with SIDM and make incorrect inferences? To this end with simulate a situation where we train on two models of CDM (CDM-fiducial and CDM-low) plus the standard two models of SIDM (0.1 and 1.0) and then see what the model predicts if we test it on the \emph{ unseen} sample of clusters from the CDM-hi simulations. The right hand panel of figure \ref{fig:combine_estimates} shows the results. We find that on estimates of the cross-section on individual estimates, the cross-section can be biased towards SIDM0.1. This is not unexpected since the higher AGN can in some ways mimic SIDM0.1. However, as we increase the sample size, this bias is reduced to $\sigma^{\rm bias}_{\rm DM}/m<0.02$. This bias originates from the fact that we infer a cross-section from a classification, such that the estimate cannot be less than zero, resulting in an implicit bias towards positive values. To mitigate this, we would carry out a full fitted regression problem, however this is beyond the scope of this paper and for future work (see section \ref{sec:disc}).

\begin{figure*}
\centering
\includegraphics[width=0.5\linewidth]{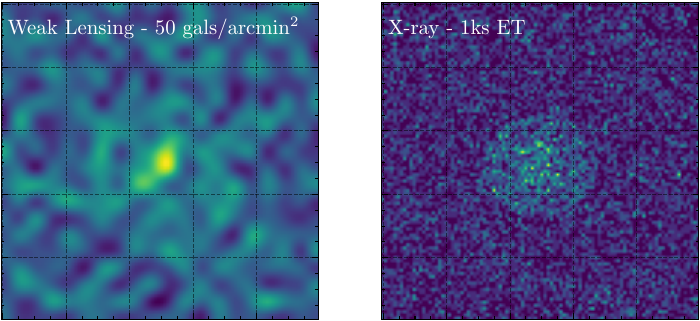}\\
\includegraphics[width=0.5\linewidth]{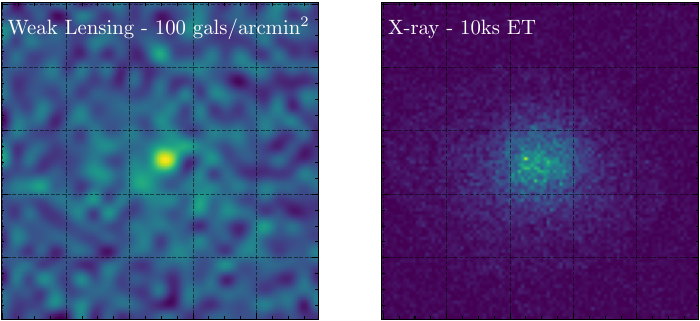}\\
\includegraphics[width=0.5\linewidth]{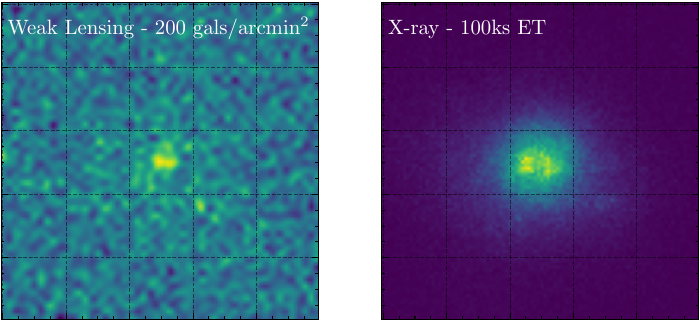}
\caption{ Observationally matched data products that are use to test the performance of the network. From top to bottom, each row represents low, middle and high signal to noise, the left column are weak lensing maps of the same cluster and  the right hand column is the X-ray emission. The top, middle and bottom row have increasing galaxy density in the weak lensing (50, 100 and 200 galaxies per square arc minute) and the right column has increasing exposure time (1ks, 10ks and 100ks exposure time (ET)).}
\label{fig:noise_ex}
\end{figure*}

\section{Making the inputs observationally realistic}\label{sec:obs}

Until now we have seen how the CNN behaves in the ideal scenario with noise-free, two-dimensional input channels. Even in the scenario where the baryonic model is unknown to us, we find that the method is robust at returning a high accuracy and separating self interacting from CDM. However, the inputs that the model is trained and tested on are not realistic. In this section we apply a pipeline that adds realistic noise to the data, in order to produce robust predictions for the performance of our model on future telescopes. An example of the noise introduced can be found in Figure \ref{fig:noise_ex} where we show the data in three regimes, low, medium and high signal-to-noise.

\subsection{Stellar Mass Maps}
\begin{figure*}
\centering
\includegraphics[width=0.51\linewidth]{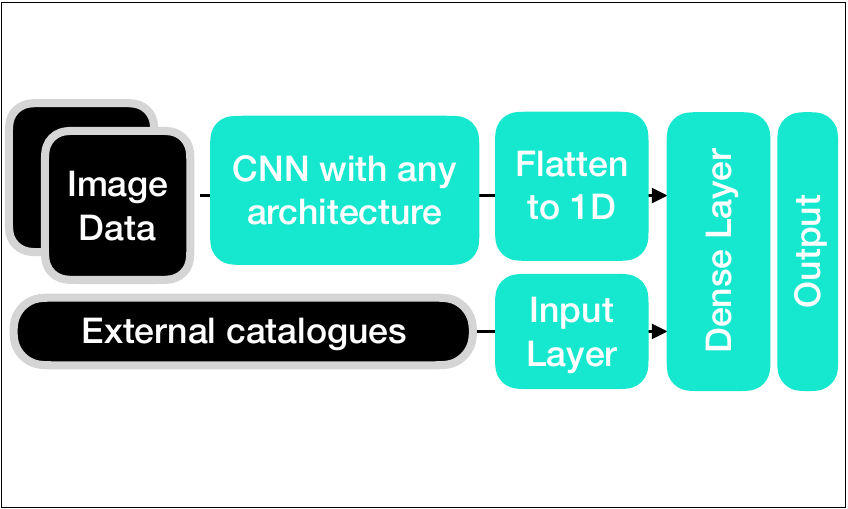}
\includegraphics[width=0.48\linewidth]{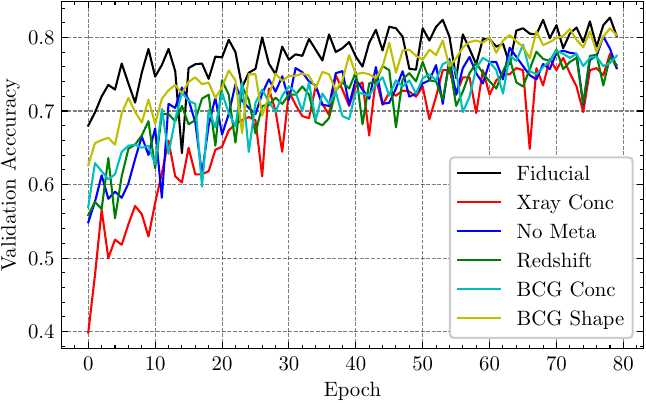}
\caption{We adapt the CNN to include an extra flattened neural network to incorporate additional external catalogues to the model. This allows us to drop the stellar mass channel and directly include a catalog of different properties. Left shows an overview of the architecture and  the right hand panel shows the impact on the accuracy of the model. We show the full, three channel model in black. We then drop the stellar channel and test different external data including the cluster x-ray emission concentration (red), no meta data (blue), cluster redshift (green), BCG mass concentration (cyan) and BCG shape (yellow). }
\label{fig:metadata}
\end{figure*}

This biggest difference between the current inputs and real data are the stellar mass maps. Indeed we use the stellar mass maps for the simulations which in real life we do not have access to. Moreover, it is non-trivial to apply the same observational noise in observed stellar mass maps to those here. To this end we developed a  network that brings together the CNN with a single layered neural network that inputs user defined meta-data. This enables the user to input single values in to the network that are combined with the image data. The architectural overview is shown in Figure \ref{fig:metadata}. We test a variety of external datasets and show the results in Figure \ref{fig:metadata}. We first show the fiducial accuracy using all three channels in black. We then drop the stellar channel and show the validation accuracy when adding different external datasets including the cluster xray-concentration (red, defined as the ratio of the enclosed x-ray surface brightness within $100$ and $400$ kpc), no meta data (however still only total and x-ray information), cluster redshift (green), Brightest Cluster Galaxy (BCG) mass concentration (cyan, the ratio of the BCG mass inside $30$ and $100$ kpc), and BCG shape, where the shape is the image moment measured ellipticity (yellow). We find that only the shape of BCG shape improves the accuracy of the model, almost fully recovering the accuracy of the fiducial model. This is extremely advantageous over using the stellar mass maps since adding noise to this input is trivial.

\subsection{ Total Mass Maps }

Gravitational lensing probes the total mass along the line of sight. Most methods to probe dark matter in clusters have used parametric methods \cite{MACSJ0416_HFF}. However, here we require non-parametric maps in-order to probe the second order effects of SIDM. We therefore choose to only use weak lensing information since this is trivial to model, albeit at the cost of losing information. To fully emulate the weak lensing process we convert the mass maps into weak lensing convergence, $\kappa$ maps via the equation:
\be
\kappa = \frac{\Sigma}{\Sigma_{\rm crit}} = \Sigma\left(\frac{4\pi G}{c^2}\frac{D_{\rm ol}D_{\rm ls}}{D_{\rm os}}\right), \label{eqn:convergence}
\ee
where $\Sigma$ is the projected surface density, and $D$ corresponds to the angular diameter distances between the observer (o), lens (l) and the source (s). We assume a lens redshift of $z_l=0.5$ and a source redshift of $z_s=1$.
We then invert the map to determine the true gravitational shear using the Kaiser Squires method \cite{KS93}. We then add on intrinsic shape noise and any additional shear measurement bias to get the observed ellipticity. Then assuming some density of source galaxies per square arc-minute, we re-invert the shear map to a convergence map, and then carry out the inference on that. The left hand column of Figure \ref{fig:noise_ex} shows three examples of simulated weak lensing mass maps with increase number density of galaxies with the top, middle and bottom row calculated with 50, 100 and 200 galaxies per square arc minute. These correspond to the expected density from large all sky surveys like Euclid \cite{EUCLID} up to the newest lensing studies with JWST where densities of 215 galaxies / square arcminute are being discovered \cite{jwst_weak_lensing}.

\begin{figure*}
\centering
\includegraphics[width=0.45\linewidth]{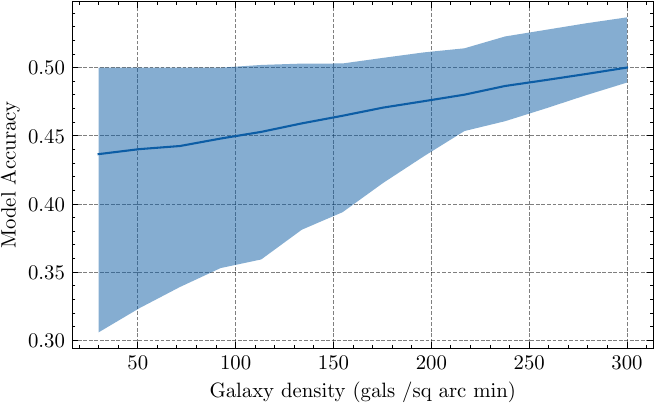}
\includegraphics[width=0.45\linewidth]{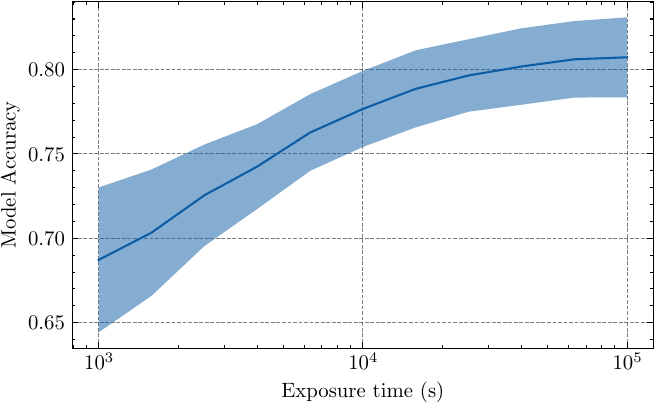}
\caption{The impact of noise on the accuracy of the neural network. We find adding weak lensing (left panel) noise significantly degrades the accuracy on a single cluster. The model is comparably insensitive to X-ray noise (right). } 
\label{fig:model_degrade}
\end{figure*}

We calculate the model accuracy degradation due to realistic weak lensing maps as a function of number density of source galaxies and show the results in the left hand panel of Figure \ref{fig:model_degrade}. We find that the model is significantly impacted by the introduction of realistic noise, with the average model reaching $45\%$ in accuracy.  However, importantly the model is still able to distinguish clusters apart, with an accuracy significantly better than random (which would be an accuracy of 33\%). This is key, since it allows us to apply an ``ensemble" method, whereby we can apply this noisy estimate to a large number of clusters and thus generate a statistical measurement of the cross-section (as long as the method is accurate).

Finally we test how systematic shape measurement bias and photometric-redshift errors can impact the accuracy of the model. No measurement of the shape of galaxies for the reconstruction of mass maps is perfect and most methods some additive ($c$) and multiplicative ($m$) bias that must be minimised in order to carry out precision cosmology, see \cite{weaklensing_rev} for a review. We explore how both a $c$ and $m$ bias impacts the accuracy of the model. We tests a range of biases from -10\% to 10\% of the input ellipticity in both $e_1$ and $e_2$. We find that the model is robust and insensitive to up to 10\% bias in the ellipticity, which is well beyond the expected shape measurement methods for future telescopes \cite{lensfitCHFT}. For the photometric redshift errors we test different source redshift distributions. We find due to the normalising of the convergence maps before they enter the CNN, the predictions are not affected, only the signal-to-noise of the maps are. Therefore we find that neither shape-measurement bias or photometric-redshift errors impact the CNN.

\subsection{X-ray Emission maps}

The X-ray maps are derived by assuming a spectral energy distribution along with the temperature, pressure and density profiles. However, these exist as noise free images. To introduce noise we mimic data from Chandra X-Ray Telescope and assume that each surface emission is Poisson sampled with a mean photon energy of 2keV at the same redshift as the lensing signal ($z_l=0.5$). We also assume a Chandra mirror diameter of 60cm. This provides an expected net count rate of photons for a given exposure time. We do not impose a point spread function on to the photos since the diffuse emission from halos are much larger than the Chandra PSF. We show in the right hand column of Figure  \ref{fig:noise_ex} the expected signal for three different exposure times of 1 kilo-seconds (top), 10ks (middle) and bottom (100ks). We calculate the model accuracy as a function of the simulated exposure time. We find, compared to the weak lensing, the model does not degrade significantly, particularly for expected exposure times of $>10ks$.

\subsection{Final predicted constraints}
\begin{figure*}
\centering
\includegraphics[width=\linewidth]{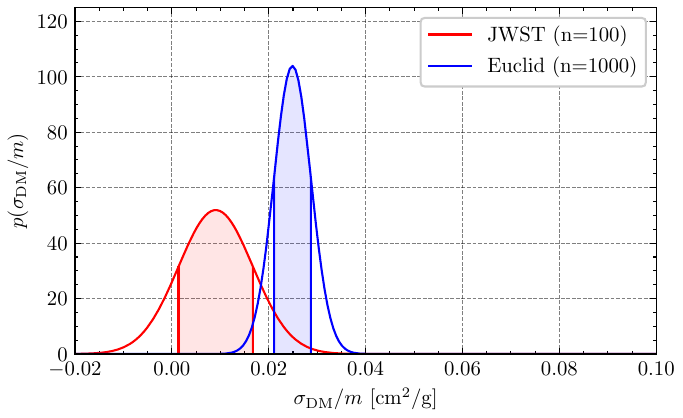}
\caption{The final constraints from a JWST style telescope (red, with deep, high resolution imaging) and a Euclid type telescope (blue). These constraints include all sources of noise in the stellar distribution, weak lensing and x-ray. } 
\label{fig:final_constraints}
\end{figure*}

In the previous section we found that the application of realistic noise degraded the accuracy of the model to $\sim45\%$. The advantages of this method is that it is quick to analyse each cluster and therefore it is logical that we develop it in to an ``ensemble method", that is, we apply our noisy but accurate method to many clusters and get a final precise constraint. This only works if each estimate is accurate and not biased. We now estimate the accuracy and constraining power of this ensemble method in two scenarios: Euclid \cite{EUCLID} which will observe more than a thousand clusters greater than $10^{14}M_\odot$ and one optimistic JWST style telescope, observing 100 clusters. We show the results in Figure \ref{fig:final_constraints}. To do this we include all sources of previously discussed noise, including:
\begin{enumerate}
\item Catalogues of the ellipticity of the Brightest Cluster Galaxy derived by measuring their image moments (c.f. how we measure in JWST imaging \cite{rrgJWST}).
    \item A finite number of clusters: 100 for JWST like and 1000 for Euclid
    \item A weak lensing map based on realistic noise estimates of either 50 (Euclid) or 100 (JWST) galaxies per square arc-minute.
    \item A constant Chandra like telescope with X-ray exposure time of 10ks.
    \item We generate predictions based on a test set that is simulated using a different seed from which the model is trained on.
\end{enumerate}

We find that with 100 JWST style clusters the ensemble method is not biased with expected constraints of $\sidm<0.02$\cpg. With an order of magnitude increase in available data (but with a degraded weak lensing precision) we would expect Euclid to deliver $<0.01$\cpg~precision. However, we see that there is a $\sim0.03$\cpg~bias caused by noise in the weak lensing estimates. This is because in the presence of noise, our model can only make estimates of the cross-section greater than zero (and not negative). Therefore any noise in the estimator will shift estimates high. This bias is very low and still out-performs any current method. This is a clear limitation of this interpolation ensemble method, however, at this sensitivity it far out performs any current method that is only sensitive to cross-sections of $\sidm>0.1$\cpg. Any future method will need more sophisticated inference to mitigate this bias.

\section{Discussion \& Future developments} \label{sec:disc}

In this paper we have presented a method to constrain self-interacting dark matter in observations of galaxy clusters using weak gravitational lensing, X-ray emission and catalogues of the Brightest Cluster Galaxy. We have carried out many tests and shown that it out-performs parametric methods that rely on centering of halos \cite{SIDM_BAHAMAS} by almost a factor of 10. During this study we carried out additional tests that showed no improvement. These can be found in \ref{app:no_improve}. Although extensively tested here, there are potential further developments that might be causes for future work:
\begin{enumerate}
    \item Regression: In this method we construct estimates of the cross-section from the probabilities from different classifications. This method results in a small but non-zero bias. A better way may be to train a regression that is able to estimate the cross-section between discretely simulated points. 
    \item Bayesian Neural Networks: We estimate the variance in the estimates by shuffling the training and test sets plus setting up mock samples of clusters and finding the combined estimate. However, this does not give estimates of true posterior of the cluster given the model. Moreover, the simplest method of combining estimates in this way also results in a small bias $\sim0.02$\cpg, which can be mitigated with a more sophisticated inference. There is a way that we can calculate the posterior for each weight in the neural net and thus output a distribution instead of a point estimator \cite{BNNarch}.
    \item Smaller cross-sections: This study shows the power of a CNN at classifying dark matter models. Indeed we are able to disentangle SIDM0.1 from different forms of baryonic feedback. In order to test this algorithm further we require smaller cross-sections and larger number of clusters. As such we will increase the training set to encapsulate both this aspects and improve it further.
    \item Finally this model is agnostic to the training data it is fed and therefore can be extended to other models of dark matter, whether it is fuzzy dark matter or other forms.
\end{enumerate}

\section{Conclusions}\label{sec:conc}

We have developed a deep learning Convolutional Neural Network that is able to distinguish between dark matter models and different levels of astrophysical feedback. We adapt an Inception model based on \cite{mertenMachine} and initially train on galaxy clusters extracted from the BAHAMAS-SIDM hydro-dynamical simulations, using 2-dimensional maps, with a field-of-view of 2Mpc binned at $20$kpc and three input channels (i.e. one sample is 100x100x3 pixels). In the idealised scenario where the input channels - total mass, stellar distribution and X-ray emission - are perfectly known, we find that our model is $\sim80\%$ accurate at classifying images of galaxy clusters in either a  CDM, SIDM0.1 ($\sidm=0.1$\cpg) or SIDM1.0 ($\sidm=0.1$\cpg) cosmology. 

We derive a method that is able to combine classifications from samples of clusters to estimate the global cross-section of dark matter. We test this on unseen SIDM0.3 data and find for individual clusters that we are unbiased with a precision of $\delta\sidm=0.1$\cpg. For a mock sample of 10 clusters we have a precision of $\delta\sidm=0.03$\cpg~ and for a sample of 100, $\delta\sidm=0.01$\cpg. We test the robustness of our method to astrophysical feedback by training the network on a separate suite of simulations that have low Active Galactic Nuclei (AGN) feedback plus the fiducial AGN feedback. We then test it on CDM with high AGN feedback, which can mimic SIDM. We find that with a sample size of 10, our method has a bias of $\sidm^{\rm bias}<0.02$\cpg. 

Finally we incorporate observational accurate inputs in order to understand how well the model performs under realistic scenarios. We extend the initial network architecture that only takes two-dimensional input data to accept parallel one dimensional external catalogues in addition to the original two-dimensional data. This enables us to drop the unrealistic stellar maps and simply include catalogues of BCG shapes. We also transform the total mass maps in to realistic weak lensing maps with a variety of signal-to-noise and finally we adapt the input X-ray maps to reflect images expected for a telescope such as Chandra. We then test on a completely new set of CDM simulations generated with a different random seed to ensure that the model has never seen the clusters before. We find including all sources of statistical noise reduces the accuracy of the model to $\sim45\%$. However, we find with the expected sample of a thousands clusters from Euclid we can reach statistical errors of $<0.01$\cpg, showing the power of this method to capitalise of the large amounts of data that will come from Euclid. 

Finally, we find that the method is completely insensitive to a host of potential systematics, including selection effects, performing equally well on both relaxed and disturbed clusters; additive and multiplicative weak lensing shape measurement bias, and photometric redshift errors, both of which are the main systematics associated with weak gravitational lensing.

Aside from these key findings, we carry out a multitude of different tests and give a brief summary here:
\begin{enumerate}
    \item Architecture: We find the more complicated Inception model out performs a simple five layer model by 10\%. We also find that pre-trained ResNet and EfficientNet do not perform well due to the required data interpolation to a large input shape.
    \item Input Channels: We find that total matter and stellar matter contribute the vast majority of the information with regard to distinguishing between SIDM and CDM, with x-ray information providing no improvement. However, X-ray is crucial in disentangling different forms of AGN feedback.
    \item Data augmentation: Adding a random rotation and flip to the training data bolsters the size of training data resulting in a over 33\% improvement in the accuracy of the model.
    \item Learning Rate: We optimise this vital meta-parameter and find a learning-rate of $10^{-3}$ is optimal.
    \item Transfer learning: We test to see if initially training on dark matter only (DMO) simulations and then fine-tuning on hydro-sims can improve the model, or reduce the requirement for large samples of hydro-simulations. We find in the limit where there is a small training set, training on DMO can improve the performance of the model by up to 5\%. However, as the training set increases the improvement decreases and becomes negligible at training sets of over 800 clusters. This should be validated when larger sample sizes are available since in the regime of large training sets there is a large scatter in the validation accuracy.
    \item Data sensitivity: We carry out a permutation importance test to see the importance of each radial bin and input channel to the performance of the CNN. We find that the importance of each pixel traces the most dominant form of matter in that pixel with central regions containing the most information. We carry out the same test on different mass bins and find that no halo mass carries more or less weight in the training sample.
    \item Baryonic confusion: We find that including x-ray information improves the ability of the CNN to disentangle different AGN feedback models by factor of three. However this induces greater confusion between CDM and SIDM. Therefore one strategy could be to use a CNN trained on different CDM models to classify the observed AGN model, then use a separate CNN trained on the classified model to then classify the dark matter model.
    \item Combining estimates: We interpret the output of the Inception model as probabilities and combine them from multiple clusters to get estimates of the global cross-section. We find for the unseen SIDM0.3 data there no bias and has an error of $\delta\sidm=0.03$\cpg~for a sample of 10 clusters and $\delta\sidm=0.01$\cpg~ for a sample of 100 clusters. We also find that our model is able to accurately measure the cross-section for a velocity dependent cross-section for a given halo mass bin.
    \item Misinterpretation of dark matter: We train on two different models of CDM plus SIDM and test it on unseen higher level of AGN feedback (as a conservative estimate) to understand whether or not our method can confuse SIDM with unseen CDM and return a false positive detection of SIDM. We conservatively estimate that our method with a sample size of at least 10 clusters has a maximum bias of $\sidm^{\rm bias}<0.02$\cpg. 
    \item Testing our algorithm on a completely new simulation generated with a different random seed degrades the performance of our method by $\sim10\%$.
    \item Adding external catalogues has a mixed impact. Adding the shape of the BCG retains almost all the information provided by the stellar maps, enabling us to drop this unrealistic channel from the training set. However, other properties such as redshift and dynamical state have no impact on the performance of the CNN.
    \item We find that our final method when validated using simulations with a new seed, all sources of statistical and systematic noise with a realistic sample size of 1000 clusters from the Euclid telescope we will be able to reach statistical errors of $<0.01$\cpg.
    \item The current method to interpret the cross-section from the simulations carries a small systematic bias of $<0.02$\cpg, which can be mitigated with more sophisticated inference methods.
\end{enumerate}

The methods we present here show how we can combine two-dimensional image data with one-dimensional catalogues to increases the sensitivity of the model but also make them more observationally applicable. This architecture can be applied to any labelled dataset and is therefore extremely useful in future studies beyond SIDM (for example fuzzy dark matter, axionic dark matter etc.).

Understanding dark matter has in the past been limited to the inference of model parameters calibrated on cosmological simulations. Deep learning presents an agnostic way to infer the model of dark matter with no assumptions on the dynamical state of the cluster or profile. Here we have shown that Convolutional Neural Networks augmented with state-of-the-art simulations have the power to deliver exciting new insights in to the nature of dark matter.

\section*{Data/Code Availability}
For all models, notebooks and information regarding the models used here, please see \url{https://github.com/davidharvey1986/darkCNN}. Data, due to its size is available on request.

\section*{Acknowledgements}
This work was supported by the Swiss State Secretariat for Education, Research and Innovation (SERI) under contract number 521107294.  We are also grateful to all those at the Pollica SIDM summer workshop and  the organisers supported by the Regione Campania, Università degli Studi di Salerno, Università degli Studi di Napoli ``Federico II", the Physics Department ``Ettore Pancini" and ``E.R. Caianiello", and Istituto Nazionale di Fisica Nucleare. DH would like to thank Andrew Robertson and Ian McCarthy in help developing and running simulations. DH would also like to thank Luca Biggio and Ethan Tregidga for insightful and helpful conversations.

\section*{Author Contributions Statement}
DH is responsible for the entire manuscript from idea conceptualisation to writing of the manuscript.

\section*{Competing Interests Statement}

There are no competing interests.

\section*{Extended Data Figures}

\newpage

\bibliographystyle{unsrt}
\bibliography{bibliography}

\newpage
\appendix
\setcounter{figure}{0}

\renewcommand{\thesection}{Supplementary Information \arabic{section}}    

\renewcommand{\thefigure}{SI~\arabic{figure}}
\section{Training Set Size}\label{app:trainingset}
\begin{figure}
\centering
\includegraphics[width=0.6\linewidth]{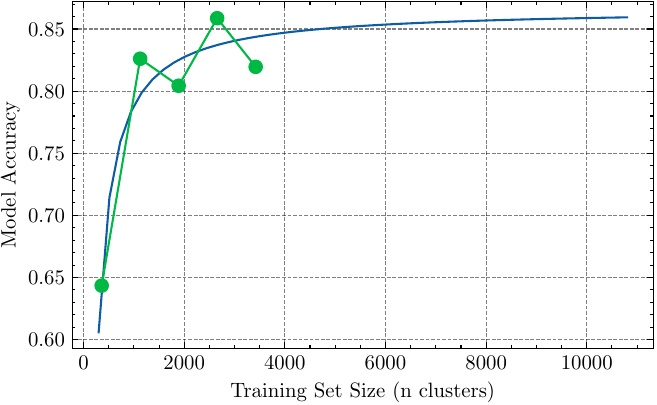}
\caption{We estimate the dependence of the model accuracy on the training set size (i.e. number of simulated clusters). By projecting clusters in to one dimension we only have 3'600 of the 10'800 clusters available. However this dramatically speeds up the training time of our model. Here we show the incremental improvement of the model with increasing training size with a fitted model in pull. We find we would only expect $\sim 2\%$ improvement if we used all available clusters. }
\label{fig:trainingsetsize}
\end{figure}

We choose to keep a small training set by only projecting clusters from the simulations to one-dimension (where in principal we could increase the training set by a factor of 3 by projecting to the other two dimensions). We do this to keep the training time to a minimum since we carry out a multitude of tests here. In future work where it is applied to real data, one could reasonably extract all clusters to get the best possible model. Here we estimate the loss of information by keeping the training set small. We take increasingly large training samples and train the Inception model with idealised three channel setup. We then test of simulations generated from a different seed to gauge the accuracy of the model (see the main text for more information regarding this). Figure \ref{fig:trainingsetsize} shows the results. We find that the model becomes relatively insensitive to the training set size  above 3000 clusters. We would expect the model to improve by about 2\% if we included the full set of clusters.

\section{Calibration of hyper-parameters}\label{app:metadata}
\begin{figure*}
\centering
\begin{minipage}{0.48\textwidth}
\includegraphics[width=\linewidth]{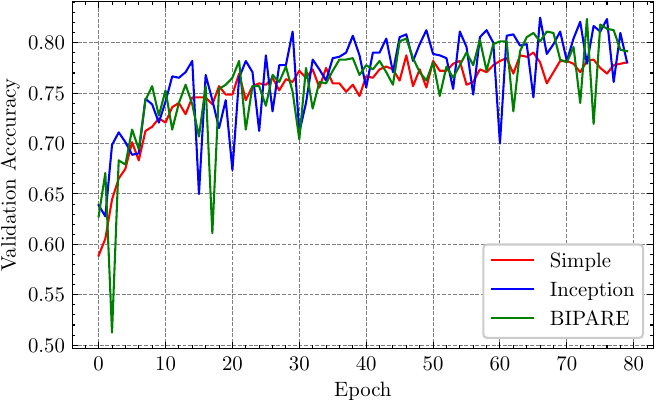}
\includegraphics[width=\linewidth]{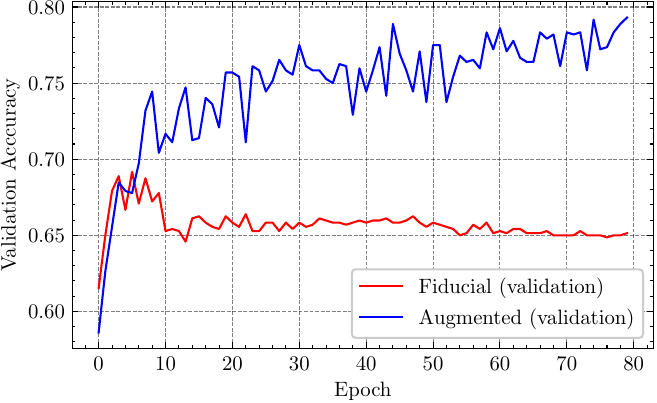}
\end{minipage}\hfill
\begin{minipage}{0.48\textwidth}
\includegraphics[width=\linewidth]{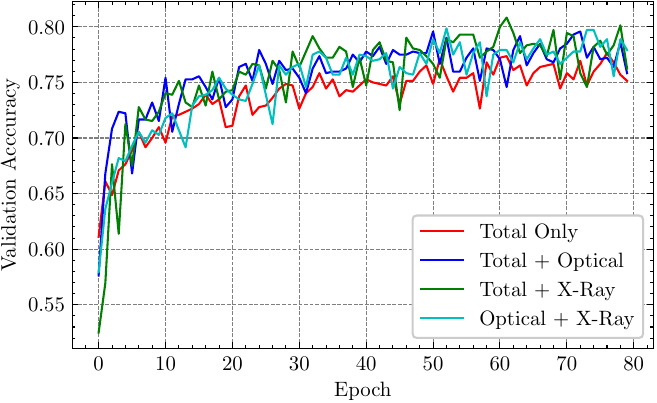}
\includegraphics[width=\linewidth]{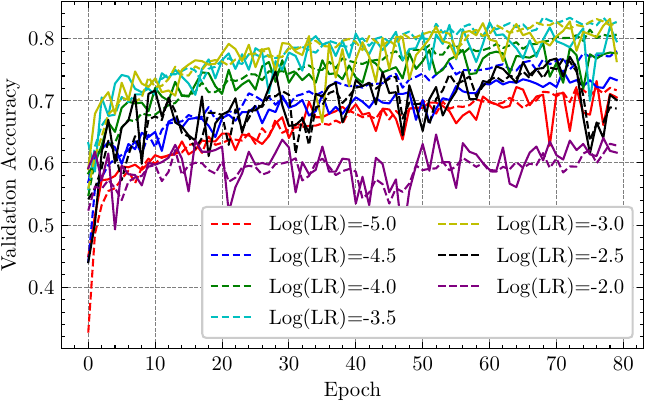}
\end{minipage}
\caption{ Four key tests associated with our CNN. In each case we show the validation accuracy as a function of the training epoch \emph{ Top Left, Architecture:} We test three models with increasing complexity. We find that the simple model (red) does not perform as well as the more complicated Inception (blue) and BIPARE (red), however the simple model does have a lower scatter and is more stable. \emph{ Bottom Left, Data Augmentation:} We show the validation accuracy both with data augmentation (blue, random rotation and random flip) and without (red). We see the significant improvement in the accuracy of the model with data augmentation. \emph{ Top Right, Input Channels:} We show the model performance for different combinations of input channels. We find that in this situation, total matter and stellar matter contribute the most information with X-ray containing almost none in deciphering between dark matter models. \emph{ Bottom right, Learning Rate:} We optimise the learning-rate (LR) meta-parameter for the Inception model. We find the best fit LR is LR=$10^-3$.}
\label{fig:tests}
\end{figure*}

As in any model there are hyper-parameters that must be tuned to garner the best performance. Here we tune the architecture, the number of channels or colours in the input image, whether to augment the data and the Learning Rate of the model, and present the results of which can be found in Figure \ref{fig:tests}. For each test we split the available data in to a training sample (80\% of the total amount of data) and a validation set (20\% of the total amount of data). We initially train on three dark matter models - Fiducial Cold Dark Matter, SIDM0.1 (i.e. $\sidm=0.1$\cpg) and SIDM1.0 (i.e. $\sidm=1.0$\cpg). We ensure for both the training and test set that there is an equal number of each class represented, and that there are equal number of clusters from each redshift slice, to ensure we do not bias the training sample. We optimise using stochastic gradient descent (ADAM) and since this is a classification problem we use the Sparse Categorical Cross-entropy as the loss function. We monitor the accuracy and validation accuracy as the main metric of success. We carry out four key tests - 
\begin{enumerate}
\item Figure \ref{fig:tests}, top left : The dependency of the validation accuracy with the architecture of the CNN. We take three example architectures: a simple CNN consisting of 5 layers and $1,203,747$ free parameters; BIPARE, a model taken from \cite{mertenMachine}, consisting of $4,185,156$ parameters and finally Inception also from \cite{mertenMachine} consisting of $13,462,756$. We find that there is a mild improvement in the accuracy between the three models. The simple model although doing much worse is significantly quicker to train, whereas the time between BIPARE and Inception is negligible despite Inception consisting of three times as many parameters. We also find that the scatter in the validation accuracy is much lower than for the Inception and BIPARE. This will be due to the low training volume spread out over larger number of free parameters. We therefore do two things. The first is we use the simple model to carry out tests on the data where architecture is not important (for example input channels and data augmentation). This speeds up the tests. Whereas any tests directly on the CNN (for example learning rate), we use the Inception model. We choose this model for one main reason, that is although the improvement is negligible and time to train is comparable to BIPARE, by choosing Inception we retain consistency between this study and \cite{mertenMachine}, which uses the Inception architecture.
\item Figure \ref{fig:tests}, bottom left : The dependency of the validation accuracy with the data augmentation. Data augmentation is a useful way to increase the size of the available training set. It was used for the classification of galaxies \cite{galaxyZooCNN} and we adopt it here. This figure shows the improvement in the validation accuracy if we include a random rotation (where we reflect the true values at the border where there exists no data) and a random flip. We find that the accuracy improves by almost $20\%$. Clearly showing the need for this augmentation. We do also test other forms of augmentation, for example, random cropping, zooming and contrast. However, these all do not improve the network. This is expected, since rotating and flipping is physically invariant, whereas cropping, contrasting and zooming actively alter the cluster, (making it less or more concentrated, changing the ratios between channels). Therefore this acts to remove information from the dataset, not enhance it.
\item Figure \ref{fig:tests}, top right : The dependency of the validation accuracy with input channel. We test to see how much information is in each channel of the CNN. Often image classification algorithms use the standard three RGB channels of images. In this case we use three classic astrophysics channels : the distribution of total matter (from weak gravitational lensing), the distribution of stellar matter (from optical imaging), and the X-ray emission maps. We find that total matter alone is able to achieve $>75\%$ in accuracy with stellar matter improving the performance by $\sim5\%$. We find adding X-ray does not significantly improve the algorithm.
\item Figure \ref{fig:tests}, bottom right: The dependency of the validation accuracy with the learning rate (LR) of Inception module. We show in this case the validation accuracy as the solid line and the dotted line the training accuracy to show that the model is not over-fitting. During back-propagation, the derivation of the Cost function with respect to the weights is calculated. At each epoch in training the weights are updated according to the calculated gradient. This ``update" is regularised by the learning rate. A large learning rate with modified the weights in the direction of the gradient in a large way, before recalculating. The advantages of this is that it can learn quickly and will not get stuck in local minima. However, it can also not converge and never find the best fitting value. Whereas a small learning rate may never reach the best fitting value and may get stuck in local minima. This parameter is therefore extremely important. We test seven learning rates, in equal log-space. We find that $\log(LR)=-3$ gives the optimal validation accuracy and will therefore use this going forward. It should be noted these tests are carried out directly on the Inception model since the LR is dependant on the number of free parameters.
\end{enumerate}

\subsection*{Increasing the sample size with transfer learning}

Given the small training set, we look at the possibility of using transfer learning to bolster the sample size. The idea behind transfer learning in this scenario is to train a model on a set of data to get estimates of the CNN weights, then to either freeze these weights and fine-tune an additional top layer to the model using a smaller training set or to fine-tune all the weights of the model. 

The advantages of transfer learning is that if the training set is computationally expensive to produce (for example simulating galaxies), then we can simulate a simpler training set for the initial model (for example a dark matter only simulation), and then fine-tune on a smaller sub-sample of the more expensive data. We carry out these test on our data. We first train our model using a suite of dark matter only (DMO) simulations. 

\begin{figure}
\centering
\includegraphics[width=0.6\linewidth]{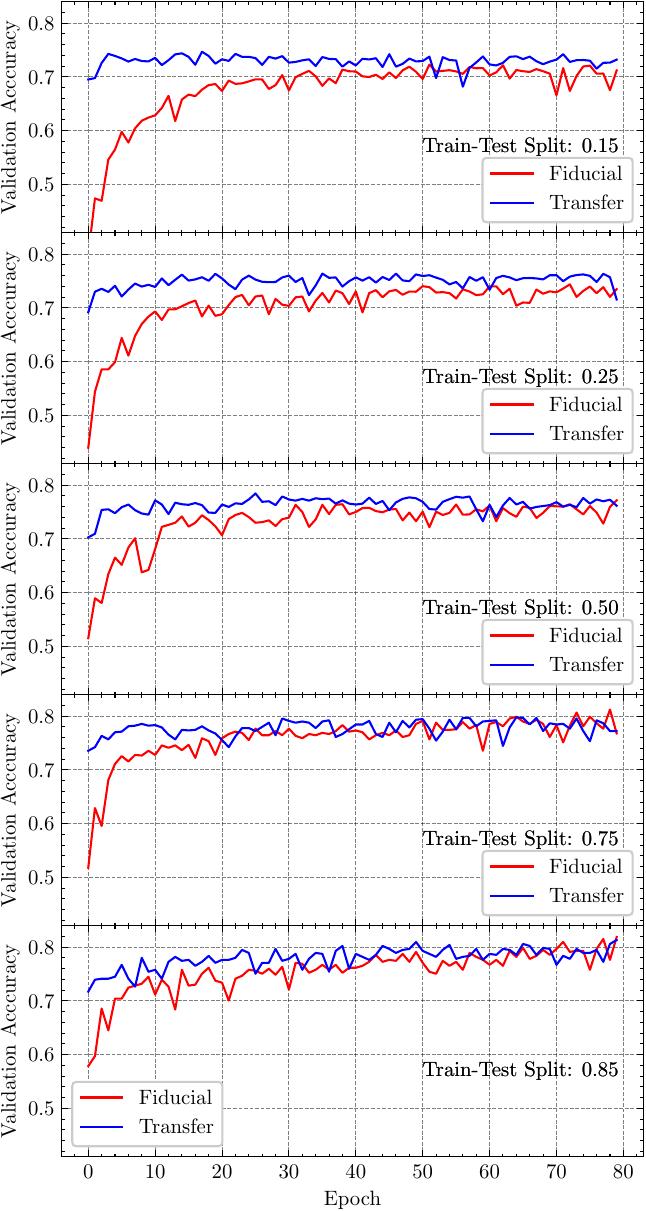}
\caption{Impact of transfer learning: We look at the improvement of initially training our model on dark matter only simulations and then fine-tuning on hydro-sims. We show the results as a function of size of training sample. We find in the limit of small training sample sizes (top panels), that initially training on DMO simulations and fine-tuning on cosmo-sims (blue line) improves the performance of the model over the fiducial training method (red line) by $\sim5\%$. However, as the training sample sizes grow (bottom panels) this improvement reduces, and becomes negligible for training sets of $>75\%$. }
\label{fig:transfer_learning}
\end{figure}

The slight complication here is that the final model will use two input channels (total matter and stellar), therefore the initial training on the DMO must also contain two input channels, which by definition is impossible since they only contain dark matter. We therefore emulate the stellar particles in the DMO with re-scaled collisionless particles. In other words, for CDM DMO we have two channels, dark matter and dark matter multiplied by the ratio of the cosmological dark matter density and the baryon density ($\Omega_M / \Omega_B$). For the interacting models, we use collisionless particles from the corresponding CDM simulation, i.e. SIDM1 dark matter plus CDM times the same ratio. This gives some estimate of the two channels. We then train these on the entire set of DMO simulations and then fine-tune them (without freezing any weights). We look at the impact of this as a function of the training set size. Figure \ref{fig:transfer_learning} shows the results. Each panel shows the impact of transfer learning for a different final training sample size. For example, the top panel has a relatively small training set (15\% of the total simulations). The red line shows the fiducial model and the blue line shows the model first trained on DMO simulations and then fine-tuned with the same training set as the fiducial model. We find for small training sizes transfer learning clearly improves the model. However at larger training sample sizes, this difference seems to disappear. It is difficult because with a larger training set, the test set is smaller and hence has more scatter. Indeed it also seems in the larger training sets that both the pre-trained model and the fiducial take the same number of epochs to reach the final accuracy, hence showing no improvement in training time. We therefore note this potentially interesting feature of training on cosmological models in low training sample sizes. However, for the training sets used here (0.85), we find no difference and therefore choose to train each model without the use of DMO simulations.

\subsection{ Tests that showed no improvement. }\label{app:no_improve}
In a bid to improve the performance of our CNN we looked in to adding different pieces of information, however in each case they did not improve the model, only added more complexity. We briefly outline them here for completeness.
\begin{enumerate}
    \item Added information - redshift and X-ray concentration. We postulated that adding information on the redshift and X-ray concentration (a proxy for the dynamical nature) of the cluster might help break some degeneracies with respect to each dark matter model. However, this did not improve the model. Indeed we find that the model does not perform better on relaxed or merging clusters when trained on the entire population. This shows that in the future observational sample selections that may preferentially select one type of cluster will not impact the parameter inference.
    \item Pre-trained models (e.g. ResNet) - we trialled taking different pre-trained models (for example ResNet \cite{resnet} and EfficientNET \cite{efficientNet}) and ``fine-tuning'' them. That is we froze the pre-trained weights of the model and simply optimised a final layer with our specific training set. We found these models exhibited significantly lower performance. This primarily came from interpolating the maps to the input shape that is required of these architecture. As such we decided to move away from these pre-trained models.
\end{enumerate}

\label{lastpage}

\end{document}